**Demonstrating the model nature of the high-temperature superconductor HgBa$_2$CuO$_{4+\delta}$**

Neven Barišić[1,2]*, Yuan Li[3], Xudong Zhao[1,4], Yong-Chan Cho[1], Guillaume Chabot-Couture[5], Guichuan Yu[3], Martin Greven[1,5]

[1] Stanford Synchrotron Radiation Laboratory, Stanford University, CA 94305, USA

[2] Institute of Physics, Bijenička 46, 10 000 Zagreb, Croatia

[3] Department of Physics, Stanford University, CA 94305, USA

[4] Jilin University, Changchun 130023, P. R. China

[5] Department of Applied Physics, Stanford University, CA 94305, USA

[*] e-mail: barisic@ifs.hr

The compound HgBa$_2$CuO$_{4+\delta}$ (Hg1201) exhibits a simple tetragonal crystal structure and the highest superconducting transition temperature ($T_c$) among all single Cu-O layer cuprates, with $T_c$ = 97 K (onset) at optimal doping. Due to a lack of sizable single crystals, experimental work on this very attractive system has been significantly limited. Thanks to a recent breakthrough in crystal growth, such crystals have now become available.[1] Here, we demonstrate that it is possible to identify suitable heat treatment conditions to systematically and uniformly tune the hole concentration of Hg1201 crystals over a wide range, from very underdoped ($T_c$ = 47 K, hole concentration p ~ 0.08) to overdoped ($T_c$ = 64 K, p ~ 0.22). We then present quantitative magnetic susceptibility and DC charge transport results that reveal the very high-quality nature of the studied crystals. Using XPS on cleaved samples, we furthermore demonstrate that it is possible to obtain large surfaces of good quality. These characterization measurements demonstrate that Hg1201 should be viewed as a model high-temperature superconductor, and they provide the foundation for extensive future experimental work.

Due to their ability to carry current without loss at relatively high temperature, high-temperature superconductors (HTSC) are increasingly used in applications such as wires[2] in current-limiting systems, power cables, motors and generators.[3] However, despite intensive efforts over the past two decades, the development of a satisfactory theoretical model of the strong electronic correlation mechanisms that give rise to the fascinating properties of these materials is still lacking. Consequently, the HTSC have presented a formidable challenge, both experimentally and in terms of developing a theoretical understanding of the measured results. Due to the complexity of their crystal structures, these materials are generally difficult to synthesize, and often only small, low-quality crystals are obtained. As is well known, the physical properties of transition metal oxides are typically very sensitive to disorder,[4] and the HTSC are no exception.[5,6] As a result, it is not always clear whether measured properties are intrinsic or sample dependent.

Among the many HTSC, $HgBa_2CuO_{4+\delta}$ is one of the most promising compounds for systematic experimental investigation due to its simple tetragonal crystal structure, its record superconducting transition temperature among all single Cu-O layer materials, and its property to confine chemical disorder to the Hg-O layers which are relatively far away from the pivotal superconducting Cu-O layers. For the above reasons, Hg1201 is a potential model system, and it is likely that quantitative experimental results obtained on well-characterized crystals will become benchmarks for tests of theoretical models. Unfortunately, Hg1201, and the Hg-based HTSC in general, have received relatively little attention so far because of the lack of sizable, high-quality single crystals. Recently, we succeeded [1] in obtaining gram-sized crystals of Hg1201 that are several orders of magnitude larger than previous samples. Here, we address several important materials-related issues associated with crystal quality such as, doping control, homogeneity, surface quality, and control over electrical contacts. These results provide valuable new quantitative information for comparison with other HTSC, and they form the basis for successful systematic future research using spectroscopic and other experimental approaches.

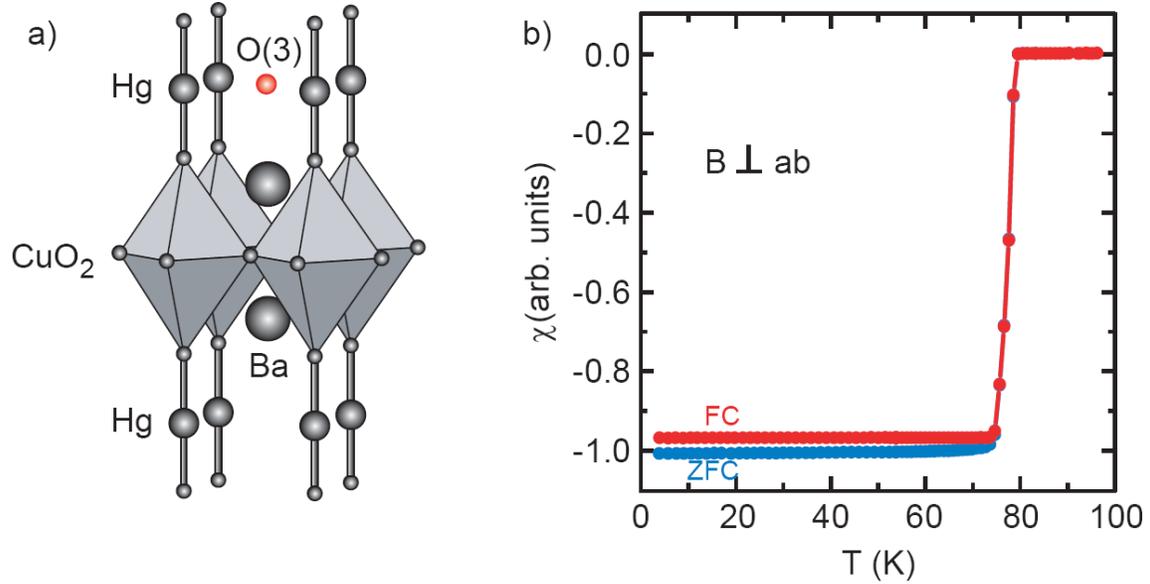

**Figure 1.** a) Tetragonal crystal structure, including the position O(3) of oxygen dopants (red circle). The Cu atoms reside at the center of $CuO_6$ octahedra. b) ZFC (measured in 10 Oe; blue) and FC (cooled in 10 Oe and measured in 10 Oe; red) susceptibility curves as a function of temperature for a sample with $T_c$ = 79(2) K (midpoint). The FC to ZFC susceptibility ratio is very high, suggesting an unusually small density of vortex pinning centers. The data shown here were obtained for a sample of mass ~ 0.3 mg which was annealed for one month at 500°C.

Hg1201 has a high-symmetry tetragonal crystal structure (P4/mmm) and a relatively simple unit cell with a small number of atoms (Fig. 1).[7] As for all HTSC, the main structural and electronic unit is the Cu-O layer, where the superconductivity is believed to originate.[8] In contrast to most other HTSC, the Cu-O layer of Hg1201 is free of long-range structural distortions. The Ba atom serves as a spacer, defining to a large extent the planar Cu-O(1) distance, and the apical oxygen O(2) resides at a significantly larger distance from the Cu atom than in other HTSC materials.[9] A special property of Hg1201 arises from the ability of the $Hg^{2+}$ cations to settle into a stable, dumbbell coordination, thus forming a strong covalent bond with the apical oxygen.[10] Consequently, there is a good lattice match between the layers and, as a result, there exists no long-range buckling of the $CuO_6$ octahedra.[11] We note that Hg1201 contains disorder in the Hg-O layer associated with excess oxygen and mercury deficiencies.[9,12] However, since this disorder is confined to the Hg-O charge reservoir layer, relatively far away from the Cu-O layers, it is generally thought not to influence appreciably the (super)conducting properties. In contrast, the quenched disorder situated immediately next to the apical oxygen atoms in many other HTSC is believed to significantly impact the local electronic properties of the Cu-O layers.[13] We note that there exists much interest in the possible existence of electronic inhomogeneities in HTSC.[14,15] Since structural and electronic inhomogeneities are closely coupled,[16,17] research on Hg1201 can be expected to provide valuable new insights, and to help differentiate intrinsic properties of the hole-doped Cu-O layers from 'extrinsic' disorder effects.

The zero-field cooled (ZFC) magnetic susceptibility is commonly used to characterize the superconducting state. The absolute value of the diamagnetic signal (Meissner-Ochsenfeld effect) together with the sharpness of the transition are usually taken as sample-quality criteria. We have measured numerous samples and generally find sharp transitions with typical widths of 2-3 K, as shown in Fig. 1b. Although it is difficult to determine the exact superconducting volume fraction due to the demagnetization factor,[18] measurements of many different samples and sample geometries point toward fully superconducting single crystals. Although the above two criteria suggest high sample quality and bulk superconductivity, they are incomplete, since they may not be sensitive to a possible spatial distribution of transition temperatures (e.g., solid and hollow spheres made from otherwise homogeneous superconducting material will have the same ZFC response). Therefore, in order to test and truly characterize the superconducting state of the crystal bulk, we also measured the field cooled (FC) susceptibility, conducted etching studies, and used remnant moment (REM) measurements [19-21] as a new way to probe sample homogeneity.

In our Hg1201 crystals, the difference between the ZFC and FC measurements (with the magnetic field perpendicular to the $CuO_2$ sheets), although sample dependent, is surprisingly low compared to other HTSC. To the best of our knowledge, the reported FC/ZFC ratio for high-quality crystals does not exceed 80%. Examples of this are $(La,Sr)_2CuO_4$ (~ 50%[22]), $YBa_2Cu_3O_{6+\delta}$ (~ 40-80%[23]), and prior work on Hg1201 (~ 30-40%[24]). The difference between ZFC and FC curves is attributed to magnetic flux trapped while cooling the sample in the field. As the temperature crosses $T_c$ upon cooling, $H_{c1}$ is low (see inset in Fig. 2b) and vortices are introduced into the bulk of the sample. By further decreasing the temperature, $H_{c1}$ increases and the vortices are expelled from the sample except for those pinned to defects. For the data shown in Fig. 1b, the FC/ZFC ratio is larger than 95%, indicating a very low density of trapped magnetic flux. This demonstrates that we were able to synthesize and select Hg1201 samples of very high quality, i.e., with a very low density of vortex pinning centers.

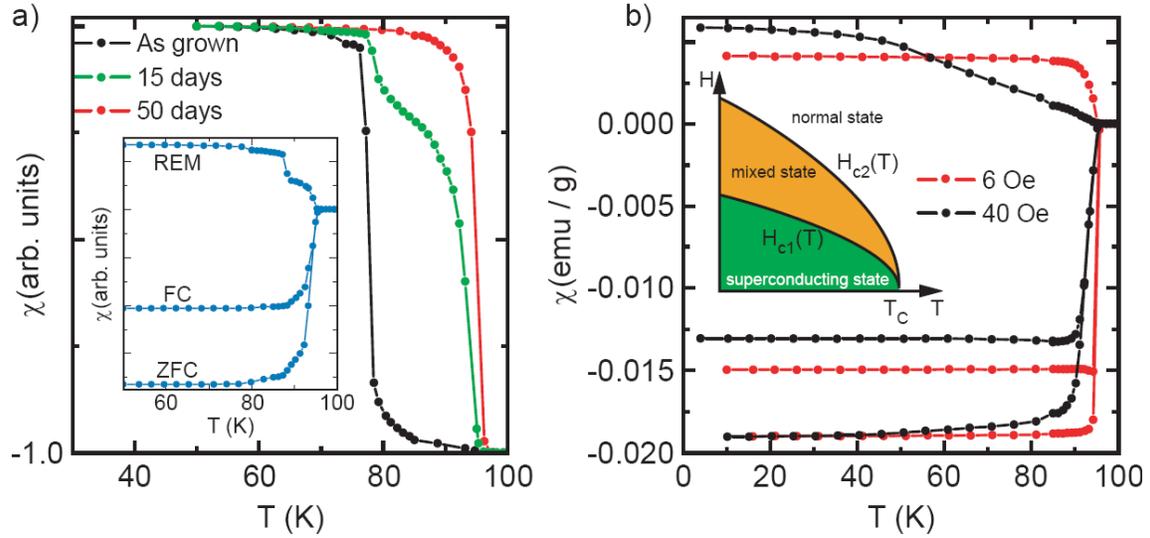

**Figure 2.** a) The REM measurement (cooled in 5 Oe) is used as a bulk probe of the superconducting state during annealing studies. For as-grown samples and for samples that were not annealed sufficiently long, the superconducting transition is typically broad and/or multi-step-like (black and green curves), indicating that the sample should be further annealed in order to change the overall oxygen concentration and/or to obtain better homogeneity. For a sufficiently-long annealed sample (red curve), the transition is sharp, which implies a macroscopically homogeneous oxygen distribution. As shown in the inset, typical ZFC and FC (measured/cooled in 5 Oe) data for an insufficiently-long annealed sample are not enough sensitive to a possibly broad distribution of transition temperatures. Those curves, although quite sharp, may suggest an incorrect value of $T_c$. We note that the FC/ZFC ratio is rather low because the measurement was performed on a larger crystal (~ 100 mg) subsequently used for an inelastic neutron scattering study. Large crystals tend to have more pinning centers. The key observation is that the REM measurement allows us to determine whether our crystals are thoroughly annealed. b) Temperature dependences of the ZFC, FC and REM magnetic susceptibility (measured moment divided by the applied magnetic field) for a homogeneous sample measured/cooled in high (40 Oe; black points) and low field (6 Oe; red points). Inset: schematic temperature dependence of the upper ($H_{c2}$) and lower critical field ($H_{c1}$) of HTSC (not to scale).[25]

A convincing way to test if there exists a $T_c$-gradient as a function of the distance from the surface is to etch the crystal and to measure its magnetic susceptibility as a function of the etched volume. We used diluted bromine acid (5% bromine, 95% alcohol) to etch a number of crystals, and established that for the majority of long-term annealed samples (two months at 350ºC in air, which resulted in $T_c$ = 95 K; typical initial sample mass of 100 mg), the temperature profile of the susceptibility did not change across the sample, indicating a rather homogeneous oxygen/disorder distribution. We note that, unless stated otherwise, we define $T_c$ as the transition midpoint. Samples that were annealed insufficiently long showed a considerably smaller and quite inhomogeneous $T_c$ deeper in the bulk, while the initial susceptibility curves reflected only the highest $T_c$ associated with the skin of the sample. In addition, it is worth noting that the initial susceptibility curves were sharp, resembling the ZFC and FC curves shown in the inset of Fig. 2a. The etching study showed that annealing times on the order of 1-2 months are sufficient to yield homogeneous samples, although crystals of different size or quality require different lengths of time.

A clear disadvantage of etching studies is their destructive nature. As a non-destructive alternative, we suggest REM measurements as a very simple bulk probe of the superconducting state. REM measurements are performed in the following way: first, the sample is cooled in a small (5-10 Oe) external field (as for FC measurements), then the field is switched off so that the diamagnetic part of the signal disappears and only the trapped flux remains, and finally the remnant moment is measured. Since the trapped flux induces superconducting currents, any macroscopic $T_c$-distribution will be easily observed: as the sample is warmed up, changes in the susceptibility correspond to those parts of the sample that are becoming non-superconducting. This is shown in Fig. 2. Although the samples were annealed for a long time and have a reasonably sharp transition as measured via FC and ZFC, the REM data shown in the inset of Fig. 2a clearly reveal several $T_c$ values (81, 87 and 96 K) and hence a broad bulk distribution. Another example of the utility of REM measurement is given in the main panel of Fig. 2a, where the effect of the oxygen inhomogeneity in the sample is followed as a function of annealing time. This experiment confirms that long-term annealed samples are homogeneously doped throughout the bulk.

For the REM measurement it is important that the applied field is sufficiently small, since otherwise additional effects can be observed. This is demonstrated in Fig. 2b, which shows ZFC, FC and REM curves measured on a high-quality sample with fields of 6 and 40 Oe (H⊥ab plane), both smaller than the low-temperature value of $H_{c1}$. In both cases, the ZFC and FC curves are flat and suggest a rather sharp $T_c$, in contrast with the REM curves. The trapped vortex density in the material increases with the applied field.

In a naïve picture, those vortices pinned to weak pinning centers will detach when the sample is heated up and find other, stronger, pinning centers. If there are no such unoccupied strong pinning centers, the vortices will be expelled from the material. The situation is more complicated if, at low temperatures, there are not enough pinning centers or if the applied field is larger than $H_{c1}$. In that case, one should also take into account the formation of a vortex lattice and its melting as the temperature is increased. For Hg1201, close to optimal doping ($T_c$ = 95 K), we established that fields of 10 Oe and smaller, when perpendicular to the ab-plane, and smaller than 1 Oe, when in the ab-plane, are sufficiently low to avoid complications of this kind.

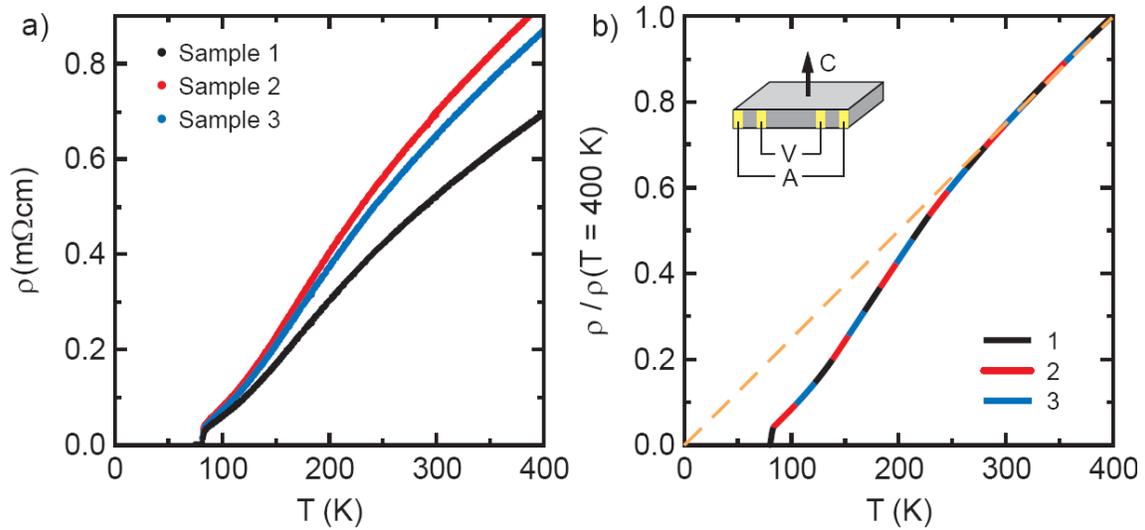

**Figure 3.** a) Resistivity for three underdoped crystals that were annealed at 520°C in air (and subsequently quenched by placing them onto a room-temperature metal plate) to yield $T_c$ = 81 K. The approximate crystal dimensions were 0.5x0.5x0.75 mm$^3$ and the absolute value of the resistivity was determined to within ~ 20%. b) The remarkable agreement of the three resistivity curves when normalized at 400 K reveals the high sample quality. Resistivity is a bulk probe that is very sensitive to disorder and the presence of secondary phases. The excellent agreement among the three curves suggests that the samples are electronically identical. The low-temperature deviation from linearity suggests that for underdoped Hg1201 with $T_c$ = 81 K the pseudogap temperature is $T^* = 250(20)$ K. The resistivity was measured using electrical leads placed as shown in the inset.

We next obtained information about the normal state from dc resistivity measurements, which, although not sensitive to small inclusions of insulating secondary phases, contain important information about the bulk. After selecting high-quality long-term annealed crystals that had been characterized by magnetic susceptibility, careful resistivity measurements were carried out. First, we used the Laue method to determine the direction of the principal crystallographic axes [1] and to ensure that the crystals exhibited clear and well-defined Bragg peaks. Next, the crystals were cleaved along the *c*-axis and four gold contacts were sputtered on the *ac/bc* faces. In this contact geometry, the measured resistivity should reflect only the electrical properties of the *ab* planes. Gold wires were then attached with silver paint and the system was baked under the same conditions as the prior long-term annealing (520°C in air, which yields a $T_c$ of 81(1) K), and then quenched to room temperature, which resulted in a contact resistance of less than one Ohm. We note that the resistive superconducting transition is sharp and coincides with that obtained from susceptibility measurements (Fig. 1). Due to the irregular sample shape, the room temperature value (approximately $\rho(300\ K)= 0.6$ mΩcm) could only be obtained to within 15-20% accuracy (Fig. 3a). As demonstrated in Fig. 3b), despite having somewhat different dimensions, the three samples display the same normalized resistivity temperature dependence, which demonstrates their high bulk homogeneity and the absence of significant disorder effects, since the latter are expected to be sample dependent. The pseudogap temperature, defined as the characteristic temperature at which the resistivity starts to deviate considerably from a linear behavior, is T* = 250(20) K, somewhat higher (by ~ 40 K) than a previous report for ceramic samples.[26] We emphasize that the low contact resistance and high reproducibility on different samples make Hg1201 exceptionally interesting for detailed future transport investigations.

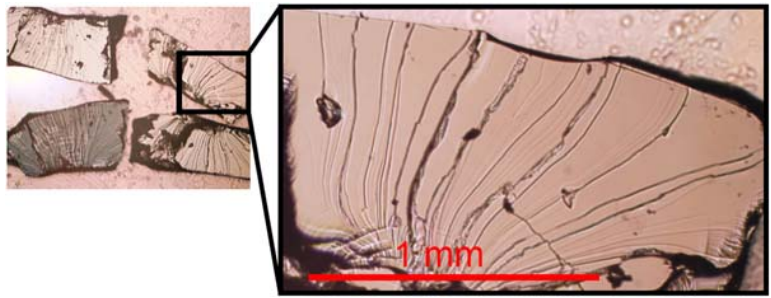

**Figure 4.** Surfaces of two cleaved Hg1201 crystals photographed by use of an optical microscope. One of the surfaces is enlarged to demonstrate its high quality.

Advanced surface-sensitive techniques such as photoemission spectroscopy and scanning tunneling microscopy provide key insights into the properties of HTSC and require uncontaminated flat surfaces obtained by *in situ* cleaving. This constraint has strongly limited the number of experimental systems that have been studied. Even for optical and Raman spectroscopy, which are less surface sensitive, it is favorable to have access to cleaved samples. Hg1201 does not exhibit a natural cleavage plane since it contains only one Hg-O layer between adjacent Cu-O layers. Nevertheless, we have been able to partially overcome this problem by using a simple modification of a standard method. As is common practice,[27] the sample is glued to a sample holder parallel to the *ab*-plane with silver epoxy and, on the opposite side, to a 'post' in order to facilitate the cleaving. We then scratched one of the four free crystal *ac*-sides with a surgical blade parallel to the *ab*-plane, which allows the crystal to cleave along the scratched line when the sample post is knocked off, in the same way that glass can be cut with a diamond blade. This technique allows the crystal to break along a chosen line, presumably along high-quality material. This is in contrast to the standard way of cleaving, where the crystal breaks at its weakest places, most probably in defect rich regions. Using this modified cleaving method, we obtained large, flat and shiny surfaces, exhibiting step-like terraces (see Fig. 4). We note that the success rate of this method is quite high (above 50%) and that it is also possible to cleave samples along the *c*-axis.

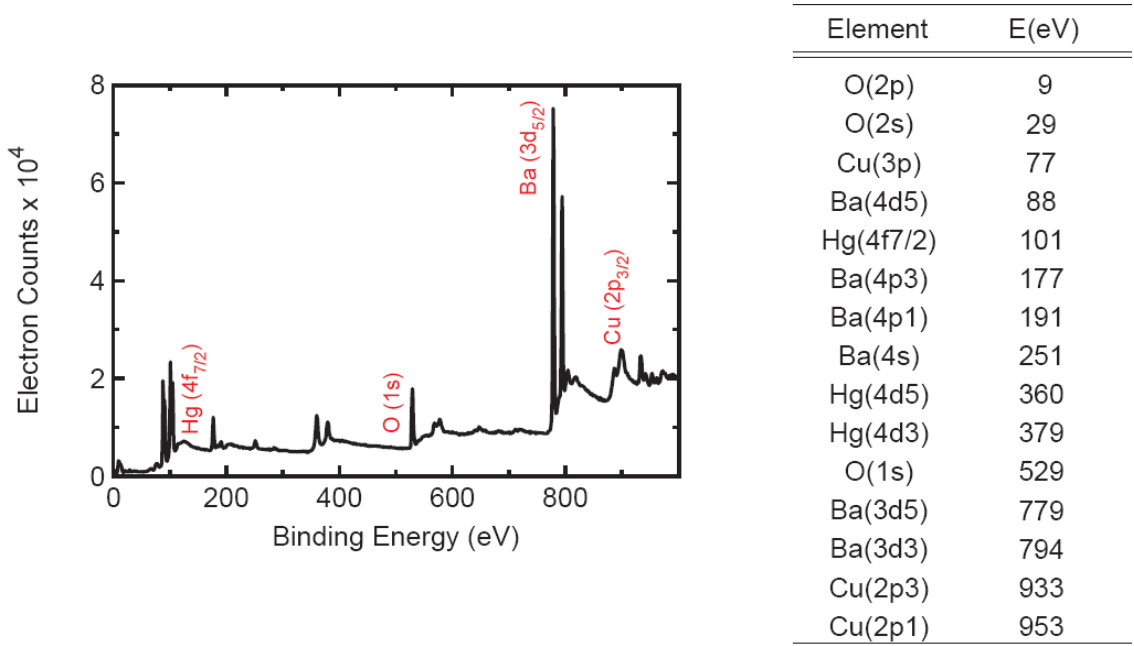

**Figure 5.** X-ray photoemission wide-energy survey scan (energy resolution: 1eV) reveals elemental composition of the surface. Within the resolution of the measurement, no trace of unwanted elements is found. All peaks are observed at the expected energy, as listed in the table.

We next performed X-ray photoemission spectroscopy (XPS) in order to check the quality of the cleaved surfaces and found the expected Hg, Ba, Cu and O peaks (see Fig. 5) for samples cleaved in vacuum (~ $10^{-6}$ Torr). We found that samples held in air for a few minutes after cleaving exhibited a strong additional carbon peak at a binding energy of ~ 285 eV (not shown). By closer examination of the oxygen O(1s) peak at 529 eV (which indicates oxygen in metal oxides), with a higher energy resolution of 0.1 eV, it is observed (not shown) that the peak is strongly suppressed after exposing the surface to air. Instead, a strong carbonate peak emerges indicating a change in the surface chemistry. We furthermore note that high-resolution XPS measurements with sufficient counting time also reveal a weak carbon peak which we believe to be a result of contamination due to the relatively low vacuum of the XPS apparatus. Consequently, when studying Hg1201 crystals with surface-sensitive techniques, one should keep in mind that the surfaces can be easily contaminated. Our new results provide a plausible explanation for the ~ 3% missing reflectivity observed in recent optical conductivity work.[28]

It is well known that electric, magnetic and structural properties of many HTSC, such as the superconducting transition temperature or the lattice constants, sensitively depend on the oxygen concentration.[29,30] The oxygen content can be varied by annealing the samples at different combinations of temperature and oxygen partial pressure. In most HTSC, this is a delicate issue since the created oxygen vacancies (or the excess of oxygen depending on the material) typically represent disorder, which in turn affects the electronic properties. In this respect, Hg1201 exhibits an advantage due to the fact that Hg forms a strong covalent bond with the apical oxygen O(2), leaving the additional oxygen, which is situated in the middle of the Hg square and rather far away from the $CuO_2$ sheets, only weakly bonded to the Hg cations (see Fig. 1).[31] Consequently, it is not expected that the extra O(3) atom appreciably perturbs the structure by introducing strain in the $CuO_2$ layer. Indeed, as discussed above (Figs. 1 and 2), the remnant field for high-quality crystals is low, which supports this hypothesis. Although the O(3) atom is not strongly bonded, the O(3) concentration defines the doping of the $CuO_2$ layers, since each such atom introduces between one [32] and two [33] holes into the superconducting planes.

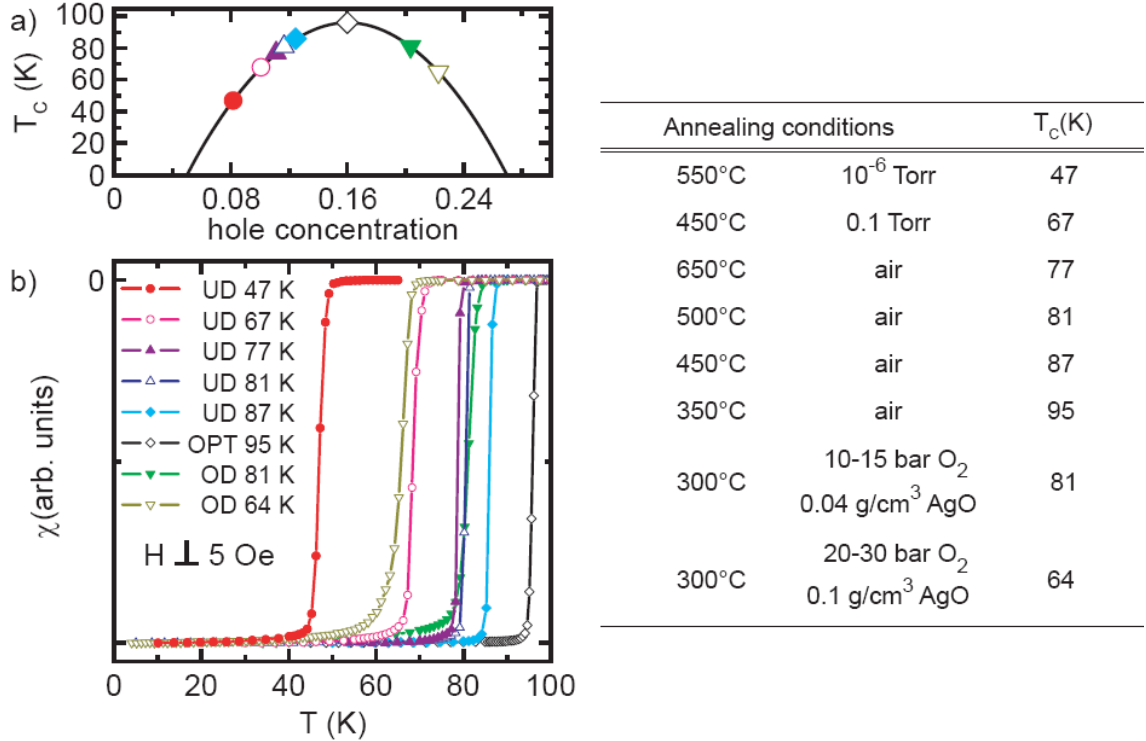

**Figure 6.** a) Based on the empirical universal equation for the dependence of $T_c$ on hole concentration, $T_c=T_{c,max}[1-82.6(p-0.16)^2]$,[34] the estimated doping range is $0.08 < p < 0.23$, which covers a large region of the phase diagram. We note that $T_c$ is defined as the transition midpoint. The highest onset transition temperature observed in our crystals was 97 K. b) ZFC temperature dependences of the magnetic susceptibility and corresponding annealing conditions (Table) for eight different samples.

We now report results of annealing studies, demonstrating an excellent control over the doping of Hg1201 for a broad range of hole concentrations. As mentioned above, oxygen homogeneity across a crystal can be achieved only through long-term heat treatment. Fig. 6 summarizes the results of our crystal annealing under many different conditions (for at least a month, depending on the sample quality and size). After the annealing, the crystals were quenched to room temperature. Because it is necessary to use relatively high oxygen partial pressures in order to overdope a crystal, we sealed several samples together with silver peroxide (the decomposition of which provides the necessary oxygen) in a quartz tube. We estimate that 0.133 g/cm$^3$ of silver peroxide increases the oxygen pressure to approximately 20 bar. In Fig. 6, ZFC curves measured with a field of 5 Oe perpendicular to the *ab*-plane are presented for five under-doped, one optimally-doped, and two over-doped samples. The superconducting transitions are found to be rather sharp (~ 2 K). REM measurements (not shown) confirmed the high quality of the samples and their homogeneous oxygen distribution after the annealing.

To summarize, our data clearly demonstrate that Hg1201 should be viewed as a model high-$T_c$ superconductor. We have succeeded in growing and selecting high-quality samples with very high FC/ZFC ratios. The bulk homogeneity of the O(3) oxygen distribution across the samples was tested through etching studies and also by measuring the remnant-magnetic-moment (REM) temperature dependence, which is a non-destructive bulk probe of the superconducting state. These findings were further confirmed by probing the normal state through careful resistivity measurements for a number of underdoped samples ($T_c$ = 81 K), all annealed under the same conditions. The reproducibility of the resistivity data is surprisingly good, and the very small value of the residual resistance suggests that the material may be essentially free of zero-temperature intrinsic spin and charge disorder.[35-37] Through extensive annealing studies, we have furthermore succeeded to dope our samples, ranging from an estimated hole concentration of 8% on the underdoped side to a concentration of about 23% on the overdoped side. It appears likely that this doping range can be further extended in the future. We also demonstrate that it is possible to obtain low-resistance electrical contacts (less than one Ohm), which is crucial for quantitative transport measurements, as well as large in-situ cleaved surfaces of good quality, which is indispensable for surface-sensitive measurements.

**Acknowledgements:** We would like to thank T.H. Geballe for helpful comments. This work was supported by the U.S. Department of Energy under Contract No. DE-AC02-76SF00515, by the US National Science Foundation under Grant No. DMR-0705086, and by Croatian Ministry of Science and Sport under Grant No. MoSES 035-0352826-2848.